# Universality of Egoless Behavior of Software Engineering Students

Pradeep Waychal, Western Michigan University, Kalamazoo, MI, USA
Luiz Fernando Capretz, Western University, London, Canada

## Abstract

Software organizations have relied on process and technology initiatives to compete in a highly globalized world. Unfortunately, that has led to little or no success. We propose that the organizations start working on people initiatives, such as inspiring egoless behavior among software developers. This paper proposes a multi-stage approach to develop egoless behavior and discusses the universality of the egoless behavior by studying cohorts from three different countries, i.e., Japan, India, and Canada. The three stages in the approach are self-assessment, peer validation, and action plan development. The paper covers the first stage of self-assssment using an instrument based on Lamont Adams' "Ten commandments (factors) of egoless programming" – seven of the factors are general, whereas three are related to coding behavior. We found traces of universality in the egoless behavior among the three cohorts such as there was no difference in egoless behaviours between Indian and Canadian cohorts and both Indian and Japanese cohorts had difficulties in behaving in egoless manner in coding activities than in general activities.

## INTRODUCTION

Software engineering concentrates much less on people than process and technology dimensions ((Broman, Sandahl, & Baker, 2012),(Dagenais, Ossher, Bellamy, Robillard, & De Vries, 2010)). Glass, et al. (2002) have studied 369 papers in six leading journals and discovered that software engineering research is fundamentally about technical and computing issues and seldom about behavioral issues. Recently, Lenberg et al. (Lenberg, Feldt, & Wallgren, 2015) noticed increased attention to the human aspects of software, but found the increase to be not sufficient. In industry, a discussion about the people dimension appears to be limited to training people for new processes and technologies (Zowghi & Nurmuliani, 2002). Since this process-technology centric approach has not accrued any perceptible gains in productivity (Brynjolfsson, 1993), we argue to explore the people dimension deeply and earnestly; even though it is new to software engineering researchers, and depends on many factors such as social and corporate ecosystems.

The paper attempts to study an important sliver of the people dimension, egoless programming, which was initially established in Weinberg's book, 'The Psychology of Computer Programming' (Weinberg, 1971). Our study introduces a multi-stage approach to develop egoless programmers. We are using contemporary terms such as egoless engineering and development and general terms such as egoless behavior to mean the same thing: egoless programming. Our multi-stage approach consists of developing an instrument to assess "egoless behavior" by individuals, validating the self-assessment with peer assessments, and formulating group and individual action plans. In this paper, we concentrate on the first stage of developing an assessment tool to gauge egoless behavior, and explore universality of the egoless behavior among software engineering students. Towards that, we have chosen three cohorts from three culturally different countries: India, Japan, and Canada. Essentially, the paper contributes to knowledge of the people dimension in software development by presenting and analyzing self-assessment of egoless behavior of students from three different countries.

The next section discusses the problem of productivity in software organizations. It is followed by the research design of our experiment. We then analyze the results and end with concluding remarks.

## BACKGROUND

Many software engineering stalwarts have emphasized the criticality of the people dimension in software engineering. Dijkstra (1979) proclaimed that programming (software engineering) has to be considered as a human activity. Weinberg (1971) clearly stated that human personality is more important than human intelligence in software. Cockburn (1999) has emphasized importance of the people dimension by stating that the fundamental characteristics of "people" have a first-order effect on software development and must become a first-order research agenda item in software engineering. Potts (1993) has claimed that "all the real problems in software engineering are people problems." Many studies have asserted criticality of teamwork in organizations (Bendifallah & Scacchi, 1989; Boehm, 1981; Mahnic, 2012; Scacchi, 1995) Therefore, the people dimension appears to be of critical importance.

It is important to note that software engineers function in groups and a greater understanding of groups from a human science perspective may help in improving group and organizational performance. This would require delving into the human science such as sociology, anthropology, organizational behavior, and psychology. However, most of the empirical software development research is performed on individual programming activities (B. Curtis et al., 1986; B. C. Curtis, 1987). Curtis and Walz (1990) asserted that software development must be studied at several behavioral levels as indicated in their layered behavioral model. The mode emphasized the factors that affect notonly cognitive, but also social and organizational processes of software development. At the individual level, only cognitive and motivational processes matter, but at the team level, social processes play a critical role. In that context, Curtis (1987) described five psychological paradigms in the realm of software development. One of the paradigms covers group dynamics, which includes team structure. Curtis has discussed two structures – centralized or chief programmer and decentralized or egoless.

Weinberg (1971) proposed an egoless structure where no central authority is invested in any specific team member. Individuals based on their relatively unique skills pick up tasks. The model, therefore, mandates a free flow of information and public ownership of all artifacts. All share the final work-product and all decisions are team decisions. In essence, the structure requires a high egoless behavior of all the team members.

Many more researchers espoused the cause of egoless behavior. Hewitt and Waltz (Hewitt & Walz, 2005) pointed out that the information system development projects require knowledge from disparate and different domains that is spread over various team members and stakeholders. They, therefore, suggested shared leadership – on the lines of the egoless programming model – to foster knowledge sharing. Faraj and Sambamurthy (2006) used two types of leaderships – directive and empowering – the latter coming close to egoless programming model. They found that empowering leadership has an important impact on team performance, especially in case of high task uncertainty or team expertise projects. Clarke, et al. (2014) proposed "in-flow peer review" – i.e., peer review done while an assignment is in progress – and underlined the importance of egoless behavior in the review process. Lewis and Smith (2008) concluded that the problem solving styles influence conflicts and performance of software engineering teams. Cockburn (1999) observed that projects progress well, when people "just talk together" and added that good project teams have to keep the person-to-person communication channels in good order. Weiss (2002) argued that documentation and programming are similar and documents developed by solo authors tend to be late, buggy, and exceedingly difficult for others to maintain. He added that egoless methods – collaborative and structured – break the proprietary connection between the writer and the artifact and facilitate stronger reviews, resulting in better quality. Acuna et al. (2009) found that the teams with the highest job satisfaction are precisely the ones whose members score highest for the personality factors, agreeableness and conscientiousness. Losada and Heaphy (2004) studied sixty teams' performances on 'other-self' dimension referring to a number of times team members refer to others versus themselves and found that the ratio was 27.5 times better in case of high performing teams as compared to low performing teams. The egoless methods are adopted by the world of free and open source software (FOSS). Eric Raymond has argued that the bazaar model – the model adopted in the FOSS world – produces better quality code than the cathedral model – the model prevalent in the commercial world (Raymond, 1999). All these studies strengthen the case for egoless behavior in software engineering.

Contemporary project teams are becoming more global. They are no longer limited to a particular locale, region, culture, or country. The Japanese business writer Keniche Ohmae aggressively says that nations have become mere fictions. While technological developments have made it possible to work across the globe, cultural differences have posed hurdles influencing success rate of contemporary projects. In this regard, Giddens (2002) argues that the era of the nation state is over. Culture plays an invisible but important role in projects. It relates to the way people think, react to events, socialize, prioritize things, and develop their work ethics. Making diverse individuals work as a single cohesive team presents a complex problem. For example, in the USA and the Netherlands, individualism is very high; whereas, in China, West Africa, and Indonesia, collectivism is very high (Olson & Olson, 2003). When individuals from such contrasting cultures undertake a project, its chances of success may be difficult to predict. Thus, the cultural aspects require serious attention and proper understanding.

The software industry seems to have overlooked this important concept in software project development. This may be due to the alien nature of human sciences, the complexity of the relationships and a lack of awareness and evidence of the impact of the human factors on software engineering. Agile methodology does emphasize more on people than process. It focuses on competency, collaboration, trust, analytics, and devolution of decision-making resulting in more person dependence (Cockburn & Highsmith, 2001). However, we have not come across studies indicating the use of human sciences while developing or using the methodology. We posit that the software engineers of tomorrow will be facing people challenges, on a larger scale and variety, and colleges have to equip students to meet those challenges.

In tune with this requirement, the Royal Academy of Engineering has identified the attributes required of graduate engineers; these include teamwork in a multi-cultural and multi-disciplinary environment (Engineering, 2007). Male and Chapman (2005) – based on the Engineers Australia accreditation board documents - point that graduate engineers must have the ability to function effectively as a leader or an individual contributor in multi-disciplinary and multi-cultural teams. The Indian National accreditation board (NBA) has developed its accreditation programs requiring similar attributes (Accrediation, 2012). The ACM IEEE joint task force has developed curricula for software engineering and mentioned that software engineers should be able to work in teams. (ACM., 2004) Thus, engineering educators – across the globe – are supporting the need to develop multi-cultural and multi-disciplinary skills among engineering students. They, further, recognize that understanding of different cultures will help significantly in developing such multi-cultural and multi-disciplinary skills.

Given such an unequivocal need expressed by the leading policy makers, many researchers have been working on cross-cultural learning. Apelian (2007) believes that one of the important skills for the 21$^{st}$ century engineer is being able to communicate, team, and understand global and current issues necessary to work effectively with people from different cultures. He adds that engineers need to understand the societal context and human aspects of their work. Erez, et al. (2013) designed an online, four-week virtual multicultural team project to test its effect on the development of cultural intelligence, global identity, and local identity of management students. The findings indicated that cultural intelligence and global identity significantly increased over time. Jiang, et al. (2012) found that the educational specialty fault line negatively predicted task-relevant information sharing, and that the nationality fault-line negatively predicted off-task social interactions that would impact group dynamics. These studies analyzed and underlined the need for working across cultures.

In summary, the people dimension has become critical for software engineering. Teaming, especially, multi-cultural teaming is its important manifestation that requires higher egoless behavior. While some ongoing research in the area has been reported, a lot requires to be done.

## RESEARCH DESIGN

The ubiquity of software is demanding better productivity and quality from software engineering. Unfortunately, software engineering has not been able to meet the demands. It is actually reporting a drop in productivity and struggling with severe quality issues. We are proposing people initiatives to complement the ongoing process and technology initiatives to tackle the problem. Software development is a team activity. Today, the teams have become more multidisciplinary and multi-national (or multi-cultural). Our experiment is laying the groundwork for improving the functioning of such multi-cultural teams by understanding egoless behavior of students from different cultures. The following sections present our approach, scope, instrument selection, data collection, reliability assessment, and data analysis.

### Approach

Our approach consists of three stages operating in a cyclic fashion: self-assessment to create awareness, team-assessment to validate the awareness, and action plan bridge the gap i.e. to develop egoless engineers. This paper covers the self-assessment stage for cohorts from three different countries. We adopted a descriptive and diagnostic type of research design. Descriptive research describes the characteristics of a population being studied and does not explore the reasons for those characteristics. Diagnostic research studies determine the frequency with which something occurs or its association with something else.

### Scope

Egoless behavior is a mindset. The earlier a mindset is developed, the better it is. Carver, et al. (2003) argue that before running an empirical study at a software company, it is useful to carry out a pilot study with students in an academic setting. Therefore, we studied engineering students from three different countries – Canada, India, and Japan. We chose the countries because we found collaborators in those countries and the countries have reasonable differences form each other. In that sense, this was a convenience sampling. The differences on key economic and social parameters of the three countries are tabulated below.

TABLE 1A : DIFFERENCES IN KEY ECONOMIC AND SOCIAL PARAMETERS IN THE THREE COUNTRIES

| Country | Canada | India | Japan |
|---|---|---|---|
| Global Corruption Index – 2013 Rank[1] | 9 | 94 | 18 |
| Ease of Doing Business – 2013 Rank [2] | 15 | 140 | 27 |
| Global Competitiveness Index 2015 Rank[3] | 15 | 71 | 6 |
| Per Capita GDP Rank[4] | 21 | 124 | 29 |
| Happiness index Rank[5] | 65 | 32 | 45 |

[1] https://www.transparency.org/cpi2013/results
[2] http://data.worldbank.org/indicator/IC.BUS.EASE.XQ
[3] http://reports.weforum.org/global-competitiveness-report-2014-2015/interactive-gci-map/
[4] http://knoema.com/sijweyg/gdp-per-capita-ranking-2015-data-and-charts
[5] https://www.gfmag.com/global-data/non-economic-data/happiest-countries

Culture refers to the way people think, feel, and act and is the result of years of evolution. It is defined as "the collective programming of the mind distinguishing the members of one group or category of people from another" (Hofstede & Hofstede, 2015) Geert Hofstede and Michael Minkov have developed Values Survey Module (VSM) for comparing culturally influenced values. The module has six values - Power Distance (large vs. small), Individualism vs. Collectivism, Masculinity vs. Femininity, Uncertainty Avoidance (strong vs. weak), Long- vs. Short-Term Orientation, and Indulgence vs. Restraint (Minkov & Hofstede, 2010).

TABLE 1B : DIFFERENCES IN VALUES IN THE THREE COUNTRIES

| Country | Canada | India | Japan |
|---|---|---|---|
| Masculine | 52 | 56 | 95 |
| Uncertainty Avoidance | 48 | 40 | 92 |
| Power distance | 39 | 77 | 54 |
| Individualistic | 80 | 48 | 46 |
| Long Term Orientation | 36 | 51 | 88 |
| Indulgence | 68 | 26 | 42 |

Power distance defines the extent to which the less powerful members expect and accept that power is distributed unequally. Individualism connotes loose ties between individuals wherein one is expected to look after oneself and one's immediate family only. Collectivism stands for integration of people into strong, cohesive in-groups right from birth. In masculine societies, men are expected to be assertive, tough, and focused on material success and women are expected to be more modest, tender, and concerned with the quality of life. In feminine society, social gender roles overlap i.e. both men and women are supposed to be modest, tender, and concerned with the quality of life. Uncertainty avoidance measures the extent to which individuals feel threatened by uncertainty, ambiguous, or unstructured situations. Long- term orientation connotes pragmatic virtues oriented towards future rewards such as saving, persistence, and adapting to changing circumstances. Short-term oriented virtues relate to the past and present such as national pride, respect for tradition, and fulfilling social obligations. Indulgence represents allowing for relatively free gratification of desires and feelings while restraint, stands for controlling such gratification. The parameters are tabulated below for the three countries (Hofstede, Hofstede, & Minkov, 2010). The higher score indicates pronounced presence of the value. The three countries appear to be different.

Hofstede was the first to bring out cultural differences across national borders. While some later studies are well appreciated, they could not surpass the overall impact of Hofstede's work. While such bipolar divisions of cultures and nation as units are being questioned in today's connected world, Hofstede's work still provides enough benefits.

## Selection of Instruments

Egoless programming as a concept is around for nearly four decades, but was not elaborated until Lamont Adams proposed ten factors called "Ten Commandments of Egoless Programming" (Adams, 2002. ). These factors, given in Table 2, seem to have found wide conceptual acceptance. We have used them to get a measure of egoless behavior and mapped the problem to the mathematical domain.

## Data Collection

We chose students, who have developed software application(s) in teams. Our cohorts came from India, Japan, and Canada, because we found collaborators in those countries. In that sense, we used purposeful sampling with an element of convenience sampling. The Indian students were from the 2$^{nd}$ semester of the junior year, the Canadian students were from the first semester of the senior year, most of the Japanese students were from the junior and senior year of the undergraduate program, and a few were from the postgraduate programs. All students' country of origin was the same as their country of study. We did not track the gender of the students, because we did not find any correlation between the gender and the egoless behavior in an Indian study. It may be there in other cultures; however, we decided to keep that outside the scope of the paper.

TABLE 2: TEN COMMANDMENTS OF EGOLESS PROGRAMMING

| Commandment | Description |
| --- | --- |
| C1 | Understand and accept that you will make mistakes. |
| C2 | You are not your code. |
| C3 | No matter how much karate you know, someone else will always know more. |
| C4 | Don't rewrite code without consultation. There is a fine line between fixing code and rewriting code. |
| C5 | Treat people who know less than you with respect, deference, and patience. |
| C6 | The only constant in the world is change. |
| C7 | The only true authority stems from knowledge, not from position. |
| C8 | Fight for what you believe, but gracefully accept defeat. |
| C9 | Don't be the guy in the room. |
| C10 | Critique code instead of people – be kind to the coder, not to the code. |

The first assessment was carried out in India, where the cohort consisted of eighty-six software engineering course students from the junior year of a computer-engineering program. The course included a semester-long software project that was developed by teams of 5-6 students each. Somewhere in the middle of the semester, a random sample of 20 students assessed themselves. They rated each factor on the Likert scale of 1 to 10 (higher the rating the higher the egoless behavior). The students were asked to indicate any ambiguity or difficulty experienced while completing the survey. After ascertaining the usability of the instrument, the entire class of 86 students assessed themselves. We explained to the students the importance of egoless behavior in their careers and use of the assessment to develop the desired behavior. We also assured them that their assessment data would not influence course grades. We received 85 valid responses; one response rated ten for all the factors and was excluded. (N1=85).

The second assessment was carried out in Japan, where the cohort consisted of 17 undergraduate and 8 graduate students of an Information Technology department. All of them had some experience in developing software. All of the students were informed about the purpose of the exercise and the criticality of being egoless in their careers. The questionnaire was translated to Japanese by a linguistics expert and reviewed by another linguistics expert. All 25 responses were valid (N2=25).

The third assessment was carried out in Canada, where the cohort consistedof senior undergraduate students majoring in software engineering. They had studied software engineering and had enrolled for a full-year capstone project course where they had to design a sizeable piece of software. The students worked on the project requirement and design in October, presented a walkthrough in November, and then started coding. The data collection took place in January 2015. All 25 responses were valid (N3=25).

All the respondents from the three countries were assured of full confidentiality of their individual inputs.

**Reliability Assessment**

It is important to conduct a thorough measurement analysis of the instrument to ensure trustworthiness of results. Test reliability indicates the extent to which individual differences in scores can be attributed to true differences. We used the most popular measure, Cronbach's Alpha, for this purpose. Table 3 shows Cronbach's Alpha values, computed using Minitab version 17, for the three countries.

TABLE 3: CRONBACH'S ALPHA VALUE - SELF ASSESSMENT

| Country | Number of Students (N) | Alpha Value |
|---|---|---|
| Indian | 85 | 0.864 |
| Japanese | 25 | 0.783 |
| Canadian | 25 | 0.854 |

Since alpha values for all the sets were found to be greater than 0.70, the instrument was judged to be reliable (Nunally & Bernstein, 1978).

**Analysis and Interpretation**

This section presents analysis of self-assessment data of the three cohorts in multiple ways. First, we analyze overall egoless behavior of the three cohorts. The instrument has three factors that pertain to coding relate behavior (C2, 4, and 10) and the remaining seven that pertain to general behavior. We analyze differences based on those two groups. After that, we analyze differences based on each factor. These three analyses depend on the scoring patterns of different cohorts,which may differ.Therefore, we analyzed differences in the factor ratings of each cohort – inter-factor ratings.

Overall egoless behavior

We ran a two-tailed p-test using Minitab v 17 on overall egoless behavior of the three cohorts. We found that that there was no significant difference between India and Canada (p value = 0.948), however, Japanese behavior was significantly lower than India (p-value = 0.0) and than Canada (0.024). We believe that the Japanese students were strict in the self-assessment; otherwise, the country, which has the best Hofstede collective value score, should have scored better. We think that the power distance value does not affect ego behavior, as the students were working in academic setting with all team members being their colleagues. The remainders of the Hofstede values, we posit, do not influence the egoless behavior.

Coding related and general egoless behavior

The factors in te instrument pertain to general egoless behavior and coding-related behavior. The factors C2, C4, and C10 relate to coding and others relate to general egoless behavior. We found no significant difference in coding related behavior between India and Canada (p value = 0.597). Like overall behavior, Japanese score was significantly lower than India (p-value = 0.0) and than Canada (0.0). The explanation provided for overall behavior also applies here.

Individual factors' behaviour

We ran the single factor ANOVA on responses to individual factors and tabulated the results in Table 4. Japanese students' egoless behavior was significantly lower on 7 factors than Indian and Canadian students (except factors 1,3,9), and Indian students' was significantly lower than Japanese and Canadian students on factor 3. We did not observe statistically significant difference in factors 1 and 9, among the three cohorts. This means that the Japanese students scored better on factor 1 (Understand and accept that you will make mistakes), factor 3 (No matter how much karate you know, someone else will always know more.) and factor 9 (Don't be the guy in the room). All these three factors represent team behavior and Japanese students seem to be much better on that. The Indian team having the lowest score on factor 3, may be due to the class comprising of highly capable students. It had almost the best 90 students from the state.

Inter-factor ratings

As the above three analyses depend on scoring patterns of different cultures, we analyzed differences in rating between different factors by each cohort by using one-way stacked ANOVA (Tukey Method) with the help of Minitab Version 17. The results are tabulated in Tables 5, 6 and 7 for Indian, Japanese, and Canadian students, respectively. As per Tukey's method, the factors that do not share a letter are significantly different.

While Japanese student responses indicated clear distinction between the factors, Canadian students showed no distinction. Indian student responses were in-between. Responses from the Japanese and Indian students indicate the coding-related factor, "You are not your code", presenting the biggest hurdle to egoless behavior. In case of Canadian students also, it has one of the lowest ratings. The coding factor indeed implies possessiveness of intellectual work by the students. Fuller and Keim (2008) quote a study of Bruns and Humphreys that describes similar traits in their wiki experiment. The general egoless behavior factors have higher ratings indicating students being egoless in their general behavior. The college environment has good camaraderie and does not have industry-like intense competition. That may have resulted in higher ratings for the generic factors.

## CONCLUSION

Software engineering has become an all-pervasive discipline. It is relied on by practically every enterprise for its programs and projects. While this engineering discipline holds promise, it is often unable to deliver the expected performance in terms of productivity, quality, and turnaround times. The challenge, we believe, requires an interdisciplinary approach. The human intensive branch of engineering needs to move beyond traditional initiatives in the processes and technology dimensions and start leveraging human sciences. That presents a number of opportunities. We have discussed one of them – egoless programming. Weinberg (1999) introduced the concept and indicated that a programming group that has conquered the ego problem can be a reality. He is proved right by the success of the open source movement that has come up with strong products such as Linux, Apache. That success has to be replicated in the commercial world.

TABLE 4: RESULTS OF ANOVA AND T-TESTS

| No | Factor | ANOVA p-value | *Ind-Jap | *Jap-Can | *Ind-Can | Interpretation |
|---|---|---|---|---|---|---|
| 1 | Understand and accept that you will make mistakes. | 0.5 | 0.85 | 0.53 | 0.24 | There is no difference in the samples from the three countries |
| 2 | You are not your code. | 0.0 | 0.00 | 0.00 | 0.59 | Japan has statistically significant lower egoless behavior with respect to both India and Canada |
| 3 | No matter how much karate you know, someone else will always know more. | 0.0 | 0.01 | 1.00 | 0.01 | India has statistically significant lower egoless behavior with respect to both Japan and Canada |
| 4 | Don't rewrite code without consultation. There is a fine line between fixing code and rewriting code. | 0.0 | 0.00 | 0.09 | 0.09 | Japan has statistically significant lower egoless behavior with respect to India |
| 5 | Treat people who know less than you with respect, deference, and patience. | 0.0 | 0.04 | 0.04 | 0.22 | Japan has statistically significant lower egoless behavior with respect to both India and Canada |

| # | Factor | | | | |
|---|---|---|---|---|---|
| 6 | The only constant in the world is change. | 0.0 | 0.00 | 0.02 | 0.30 | Japan has statistically significant lower egoless behavior with respect to both India and Canada |
| 7 | The only true authority stems from knowledge, not from position. | 0.0 | 0.00 | 0.01 | 0.88 | Japan has statistically significant lower egoless behavior with respect to both India and Canada |
| 8 | Fight for what you believe, but gracefully accept defeat. | 0.0 | 0.00 | 0.06 | 0.76 | Japan has statistically significant lower egoless behavior with respect to both India and Canada |
| 9 | Don't be the guy in the room. | 0.22 | 0.25 | 0.14 | 0.28 | There is no difference in the samples from the three countries |
| 10 | Critique code instead of people – be kind to the coder, not to the code. | 0.00 | 0.00 | 0.00 | 0.08 | Japan has statistically significant lower egoless behavior with respect to both India and Canada |

- The columns indicate p values for t-tests between students from those two countries.

TABLE 5: GROUPING INFORMATION USING TUKEY METHOD – SELF-ASSESSMENT OF INDIAN STUDENTS (N=85, SCALE 1-10, 10 BEING THE MOST EGOLESS BEHAVIOR)

| Factor | Mean | Group |
|---|---|---|
| The only true authority stems from knowledge not from position | 8.23 | A |
| Treat people who know less than you with respect and patience | 8.19 | A |
| No matter how much karate you know someone else will always know more | 8.16 | AB |
| Fight for what you believe but gracefully accept defeat | 8.10 | AB |
| Understand and accept that you will make mistakes | 8.01 | AB |
| Critique code instead of people – be kind to the coder not to the code | 7.99 | AB |
| Don't be the guy in the room. | 7.78 | AB |
| Don't rewrite code without consultation | 7.63 | AB |
| The only constant in the world is change | 7.62 | AB |
| You are not your code | 7.51 | B |

TABLE 6: GROUPING INFORMATION USING TUKEY METHOD – SELF-ASSESSMENT OF JAPANESE STUDENTS
(N=25, SCALE 1-10, 10 BEING THE MOST EGOLESS BEHAVIOR)

| Factor | Mean | Group |
|---|---|---|
| No matter how much karate you know, someone else will always know more | 9.16 | A |
| Understand and accept that you will make mistakes | 8.04 | AB |
| Treat people who know less than you with respect and patience | 7.40 | ABC |
| Don't be the guy in the room. | 7.24 | ABCD |
| Fight for what you believe, but gracefully accept defeat | 7.12 | BCD |
| The only true authority stems from knowledge, not from position | 6.60 | BCDE |
| Critique code instead of people – be kind to the coder, not to the code | 6.20 | BCDE |
| Don't rewrite code without consultation | 5.96 | CDE |
| The only constant in the world is change | 5.44 | DE |
| You are not your code | 5.12 | E |

TABLE 7: GROUPING INFORMATION USING TUKEY METHOD – SELF-ASSESSMENT OF CANADIAN STUDENTS
(N=25, SCALE 1-10, 10 BEING THE MOST EGOLESS BEHAVIOR)

| Factor | Mean | Group |
|---|---|---|
| Treat people who know less than you with respect and patience | 8.64 | A |
| No matter how much karate you know, someone else will always know more | 8.61 | A |
| Critique code instead of people – be kind to the coder, not to the code | 8.54 | A |
| Understand and accept that you will make mistakes | 8.36 | A |
| The only true authority stems from knowledge, not from position | 8.18 | A |
| Don't be the guy in the room | 8.11 | A |
| Fight for what you believe, but gracefully accept defeat | 8.07 | A |
| You are not your code | 7.68 | A |
| The only constant in the world is change | 7.18 | A |
| Don't rewrite code without consultation | 7.00 | A |

We have proposed a multi-stage approach for developing egoless software engineers and have analyzed the first stage of self-assessment. Our experiment in three different countries showed some common traits in the egoless space, despite marked differences in their socio-economic and cultural backgrounds. The Indian and Canadian students' ratings did not have statistical differences. The coding related behavior in both Indian and Japanese students presented more hurdles to egoless behavior. We also found some differences in the responses. Japanese students had significant differentiation in response to the ten factors, Indians had some differentiation, and Canadian students did not have any differentiations. The single factor ANOVA was run on all ten factors. It indicated two factors had no statistically significant difference amongst three countries; seven of the factors have Japanese students' responses lower and one factor having Indian students' responses lower. Overall, lower Japanese responses may be due to the higher Japanese standards of teamwork. The three questions where the Japanese students scored better indicate their strong democratic work culture.

We are not presenting our findings as the conclusive evidence but only a possibility. They require to be reinforced with more such experiments. We also have to extend the research to the next steps of team assessment to validate the self-assessment and development of action plans to improve the behavior. Assessment data of individual team members can be aggregated to team's egoless index. We have to confirm the correlation and causality between such team indices and project performances, first in academic setting and then in industry setting. We need to devise and execute development plans based on the assessments and check their impact on the team indices. We also need to expand the experiment to different settings – including geographical areas and various types of software houses – and validate the findings. Owing to different team dynamics, which are based on many factors such as the project at hand, team members, and organizational cultures, individual measurements will need to be carried out in many different projects, even with the same sample, to increase their credibility.Further, the study may be applied to pertinent activities of other engineering branches. We believe that the work done so far brings out an interesting possibility of the universality of egoless behaviors, and has utility to practitioners, educators, and researchers. It can open avenues for further research in team compositions and people dynamics in project organizations to maximize their performances. The paper assumed nationality or nation state as the basic unit of analysis. As individuals participate in multi-cultural projects, the boundaries of the basic units will break blurring some of the cultural differences. It will be interesting to see which factors get impacted in such cross-cultural milieus and which do not.

## ACKNOWLEDGEMENTS

We thank all the students and who participated in the exercise and our Japanese associates Ms. Ayano Ohsaki and Dr. Niimura who facilitated data collection from the Japanese cohort. We also thank Mr. Abhay Joshi who improved the language of the paper and anonymous reviewers who, with their insightful comments, helped us improve the contents of the paper.

*Pradeep Waychal has 32 years of experience in renowned business and academic organizations such as Patni computer systems, NMIMS University, College of Engineering, Pune, IIT Bombay and IIT Delhi. Pradeep has published papers in peer reviewed journals, presented keynote/invited talks in many high profile international conferences. He/his teams have won a range of awards in Engineering Education, Six Sigma and Knowledge Management at international events. He has completed Ph.D. in the area of Information Technology and Innovation Management from IIT Bombay and is pursuing research in engineering education, software engineering and innovation management. He is currently working as a visiting professor at Western Michigan University. He can be contacted at pradeep.waychal@wmich.edu*

*Luiz Fernando Capretz is a professor of software engineering and assistant dean (IT & e-Learning) at Western University in Canada, where he also directed a fully accredited software engineering program. He has vast experience in the engineering of software and is a licensed professional engineer in Ontario. Contact him at lcapretz@uwo.ca or via www.eng.uwo.ca/people/lcapretz/ .*